\documentclass[%
 reprint,superscriptaddress,
 amsmath,amssymb,
 aps,prb
]{revtex4-2}

\usepackage{graphicx}
\usepackage{dcolumn}
\usepackage{bm}
\usepackage{braket}
\usepackage{multirow}

\newcommand{\calL}{{\cal L}}
\newcommand{\calK}{\mathcal{K}}

\newcommand{\calO}{\mathcal{O}}

\begin{document}

\preprint{APS/123-QED}

\title{
Resolving the Berezinskii-Kosterlitz-Thouless transition in the 2D XY model \\
with tensor-network based level spectroscopy
}

\author{Atsushi Ueda}
\affiliation{%
 Institute for Solid State Physics, University of Tokyo, Kashiwa 277-8581, Japan
}%
\author{Masaki Oshikawa}
\affiliation{%
 Institute for Solid State Physics, University of Tokyo, Kashiwa 277-8581, Japan
}%
\affiliation{Kavli Institute for the Physics and Mathematics of the Universe (WPI),
The University of Tokyo, Kashiwa, Chiba 277-8583, Japan}
\affiliation{Trans-scale Quantum Science Institute, University of Tokyo, Bunkyo-ku, Tokyo 113-0033, Japan}

\date{\today}

\begin{abstract}
Berezinskii-Kosterlitz-Thouless transition of the classical XY model is re-investigated, combining the Tensor Network Renormalization (TNR) and the Level Spectroscopy method based on the finite-size scaling of the Conformal Field Theory.
By systematically analyzing the spectrum of the transfer matrix of the systems of various moderate sizes which can be accurately handled with a finite bond dimension, we determine the critical point removing the logarithmic corrections. This improves the accuracy by an order of magnitude over previous studies including those utilizing TNR.
Our analysis also gives a visualization of the celebrated Kosterlitz Renormalization Group flow based on the numerical data.
\end{abstract}

\maketitle

\section{Introduction}
The Berezinskii-Kosterlitz-Thouless (BKT) transition was historically the first example of topological phase transitions, which is now an essential concept in physics~\cite{Nobel2016}.
The canonical model exhibiting the BKT transition is the classical XY model in two dimensions, which is defined by the total energy (classical Hamiltonian)
\begin{eqnarray}
\mathcal{E} =-\sum_{\langle{i,j}\rangle}\cos(\theta_i-\theta_j)\label{definition_XY},\label{XY_spin}
\end{eqnarray}
where $\theta$ takes the angular value $(0\leq\theta<2\pi)$ at each site on the square lattice and only the nearest-neighbor interactions are considered.
The topology of the configuration space allows topological point defects (vortices and antivortices), whose dissociation drives the BKT transition.
The BKT transition can be described in terms of two running coupling constants:
$y_\calK$ controlling the thermal fluctuation of spin waves, and $y_V$ representing the vortex fugacity
(see Eq.~\eqref{XXZ_Hamiltonian_appendix} for the precise definitions).

The nontrivial interplay between these two couplings is described by 
the celebrated Kosterlitz renormalization group (RG) equation~\cite{kosterlitz1974critical}
\begin{equation}
    \begin{aligned}
    \frac{dy_\calK}{dl}&=-y_V^2 , 
\\
    \frac{dy_V}{dl}&=-y_Vy_\calK ,
\end{aligned}
    \label{RG_BKT}
\end{equation}
where $l \sim \log{L}$ is the logarithm of the length scale $L$~\footnote{Including the higher order terms in $y_\calK$ and $y_V$, the
right-hand side of the RG equation may be modified. However, its form depends on the choice of the coupling constants (renormalization prescription), although measurable quantities should be independent of the choice.
Following Ref.~\cite{lukyanov1998low}, we adopt the prescription such that Eq.~\eqref{RG_BKT} is exact to all orders in $y_\calK$ and $y_V$.}.

\begin{figure}[tb]
 \begin{center}
  \includegraphics[width=86mm]{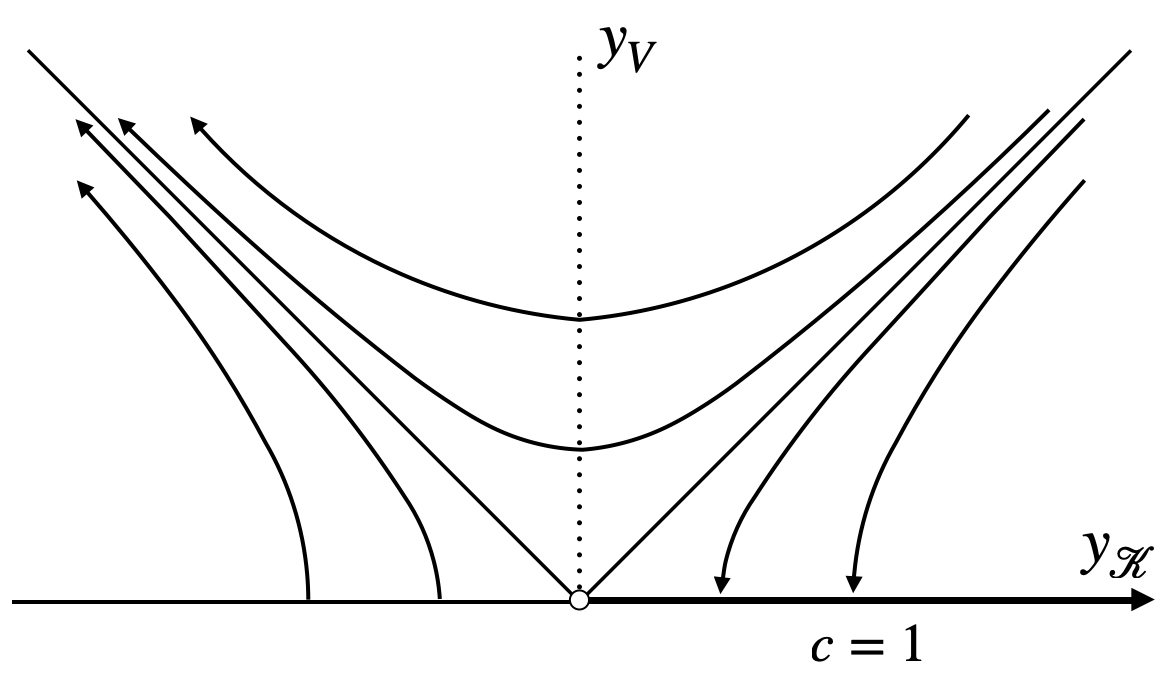}
 \end{center}
 \caption{Phase diagram of the BKT transition of the XY model. The behavior of the low-temperature phase is controlled by the $c=1$ critical line originated from U(1) symmetry. The RG flow separates the phase with $y_V\rightarrow0$ from $y_V\rightarrow\infty$ phase when $l$ increases}
 \label{KTphase}
\end{figure}

The phase diagram is Fig.\ref{KTphase} and the transition line becomes $y_\calK=y_V$.
We can introduce new variables $g$ and $t$ by $y_V = g + t$ and $y_\calK = g -t$, so that the phase boundary corresponds to $t=0$.
At the BKT transition, $t=0$, the RG equation for $g$ is reduced to $dg/dl = - g^2$,
which implies $g \sim 1/l \sim 1/\log{L}$. This slow decay is the source of the notorious logarithmic corrections.

The BKT transition is conceptually well understood in terms of the Kosterlitz RG equation.
However, the famous ``Kosterlitz RG flow'' has remained a rather abstract concept which cannot be seen directly.
Moreover, because of the logarithmic corrections, significant finite-size effects persist even in a large system, making conventional finite-size scaling of Monte-Carlo methods such as Binder plot~\cite{PhysRevB.30.3980,EDWARDS1991289} powerless. Even with considerable efforts over decades, an accurate determination of the critical temperature remains difficult even with a huge computational power.

On the other hand, many 1D quantum systems can be also described by the same effective theory and thus also exhibit the BKT transition.
Interestingly, a powerful numerical finite-size scaling method called ``Level Spectrscopy'' was developed specifically for those 1D quantum systems by Okamoto and Nomura~\cite{nomura1995correlation,OKAMOTO1992433,nomura1994critical,nomura1998symmetry,matsuo2006berezinskii}.
Based on the Conformal Field Theory (CFT) results on the finite-size energy spectrum~\cite{cardy1984conformal,cardy1986operator}, they found that the BKT transition can be identified with a level crossing between a certain pair of the energy levels, canceling the logarithmic corrections.
This allows a surprisingly accurate determination of the BKT transition point with exact numerical diagonalization of rather small systems.

\par{}
However, the applications of Level Spectroscopy have been limited to 1D quantum BKT system such as quantum spin chains so far.
While it should be applicable to the spectrum of the transfer matrix for the classical 2D XY model, the Level Spectroscopy has not been applied there, because of the difficulty in calculating the spectrum for the system with continuous variables $\theta_j$. 
In this paper, we demonstrate a successful implementation of Level Spectroscopy on the classical 2D XY model, based on the Tensor Network Renormalization (TNR) scheme~\cite{PhysRevLett.99.120601,PhysRevLett.115.180405,PhysRevB.95.045117,PhysRevLett.118.110504,PhysRevLett.118.250602,PhysRevB.97.045111}.
The TNR enables to obtain a precise spectrum of the transfer matrix, to which the Level Spectroscopy can be applied.
As in the case of the 1D quantum system, it allows a very accurate determination of the critical point by removing the logarithmic corrections from systems of moderate sizes which can be described by a tensor network with a finite bond dimension.
On the other hand, our TNR study covers larger system sizes than those in the existing Level Spectroscopy studies on 1D quantum systems.
We find a new feature of the finite-size scaling, that leads to a further improvement of Level Spectroscopy.
Moreover, we can also visualize the celebrated Kosterlitz RG flow of the BKT transition from numerical data, for the first time to our knowledge~\footnote{As a related work, in Ref.~\cite{Zou_2018}, the RG flow from the tricritical Ising to the Ising fixed points in a 1D quantum system was computed based on the Matrix Product State. The study on the RG flow of the sine-Gordon model by a nonperturbative functional renormalization-group
approach was done in Ref.~\cite{PhysRevLett.122.155301}.}
\begin{figure}[tb]
    \centering
    \includegraphics[width=86mm]{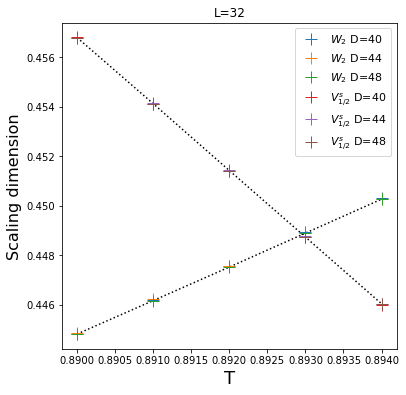}
    \caption{The calculated scaling dimension of $x_{W_2}$ (increase as temperature goes up) and $x_{V^s_{1/2}}$(decrease as temperature goes up) using Loop-TNR\cite{PhysRevLett.118.110504} at $L=32$. As $x_{W_{\pm2}}$ are always degenerate, the level-crossing with $x_{V^s_{1/2}}$ implies the formation of the SU(2) triplets.  The calculation is carried out with various bond dimension $D=$40, 44 and 48 . The level-crossing points are obtained by linearly fitting two lines. The data points “+” are on top of each other.}
    \label{level_crossing}
\end{figure}
\section{SU(2) symmetry on the BKT line}
The low-temperature critical phase of the BKT transition is described by the free boson field theory in $1+1$ dimensions, also known as Tomonaga-Luttinger Liquid:
\cite{kadanoff1979multicritical,tomonaga1950remarks,luttinger1963exactly}.
\begin{align}
 H^{\text{TLL}} = \int dx\left[\frac{\pi K}{2}\Pi^2+\frac{1}{2\pi K}(\partial_x \phi)^2\right] ,
 \label{Gaussian_appendix}
\end{align}
where $K$ is called the Luttinger parameter and $\phi$ is a dual field of $\theta$  with compactifications $\phi \sim \phi + \pi$ and $\theta \sim \theta + 2\pi$. $\Pi=\frac{1}{\pi K}\dot{\phi}$ is the canonical conjugate field of $\phi$. Note that we deal with a continuous field $\theta(x)$ instead of the lattice variables $\theta_i$. (We follow the convention of Ref.~\cite{giamarchi2003quantum}. See also the Appendix.)
The compactification of the fields entails the existence of vertex operators $V_{m,n}=\ :e^{im\theta+i2n\phi}:$, which have the conformal weights $h_{m,n}=\frac{1}{2}(\frac{m}{2\sqrt{K}}+\sqrt{K}n)^2$ and $\bar{h}_{m,n}=\frac{1}{2}(\frac{m}{2\sqrt{K}}-\sqrt{K}n)^2$.

It is helpful to first consider the BKT transition in the quantum spin-$\frac{1}{2}$ XXZ chain.
There, the single vertex operators $V_{0,\pm 1}$ are forbidden~\cite{Haldane-preprint,Affleck-LesHouches}
and the BKT transition is driven by the double vertex operators $V_{0,2}+V_{0,-2}$.
The Hamiltonian is then modified to the sine-Gordon model as,
\begin{align}
    H_{\text{XXZ}}=H^{\text{TLL}}_{K=1/2}+\int\frac{dx}{2\pi}\left\{\frac{y_\calK}{2} \calK+{y_V}\cos(4\phi)\right\} + H',
    \label{XXZ_Hamiltonian}
\end{align}
where $\calK=-2\partial_\mu\phi\partial^\mu\phi$ shifts the Luttinger parameter from $K=\frac{1}{2}$ and $H'$ represents the RG-irrelevant perturbations. 
The RG equation is again Eq.~\eqref{RG_BKT} and $g=y_\calK=y_V$ is the BKT transition line. This corresponds to the isotropic XXX chain and the system has SU(2) symmetry. More precisely, $H^{\text{TLL}}_{K=1/2}$ turns into SU(2)$_1$ WZW model and Eq.~\eqref{XXZ_Hamiltonian} can be written with SU(2) currents as~\cite{PhysRevLett.55.1355,PhysRevB.55.8295} (See Sec.~A and Table~II of the Appendix for details)
\begin{align}
    H_{\text{XXZ}}
    =  & H^{\text{WZW}}_{k=1}-  \int\frac{dx}{2\pi} \left[ g  (\bm{J^L}\cdot \bm{J^R}) \right.
\notag \\ & 
\left. 
    + t \left( \frac{J^+\bar{J}^-+J^-\bar{J}^+}{2} - J^0 \bar{J}^0 \right) 
    \right] + H' .
\end{align}
Ignoring the RG-irrelevant perturbation $H'$ for the moment, exactly at the BKT transition, $t=0$ and the effective Hamiltonian has a SU(2) symmetry.
Consequently, all the finite-size energy levels can be classified in terms of the representation of the SU(2).
Therefore, the BKT transition point can be identified by the degeneracy (level crossing) of the finite-size energy levels.
This is nothing but the Level Spectroscopy method proposed and developed in Refs.~\cite{nomura1995correlation,OKAMOTO1992433,nomura1994critical,nomura1998symmetry,matsuo2006berezinskii,PhysRevB.86.024403,PhysRevB.91.045121,PhysRevLett.115.080601}.
Thanks to the emergent SU(2) symmetry, the transition point determined by the Level Spectroscopy is \emph{exact in all orders of the marginal coupling $g$}, as long as the irrelevant perturbation $H'$ is negligible.
In CFT, there is a correspondence between the finite-size energy levels and local fields.
The lowest excited states of $H^{\text{WZW}}_{k=1}$ correspond to the ``spin-wave'' operators $W_{\pm 1} = e^{\pm i \theta}$ and the single vortex
operators $V_{\pm 1} = e^{\pm 2 i \phi}$, all with the scaling dimension $1/2$.
These $4$ states are split into a singlet and a triplet by the SU(2) symmetric marginal perturbation $g$.

In the 2D classical XY model, the single vortex operator $V_{\pm 1} = e^{\pm 2 i \phi}$
is not forbidden in the Hamiltonian and indeed it is what drives the BKT transition, rather than the double vortex operator $V_{\pm 2}$.
Nevertheless, the effective field theory in terms of boson field is equivalent to Eq.~\eqref{XXZ_Hamiltonian}, with the replacement $2\phi\rightarrow\phi$ and $\theta\rightarrow2\theta$. This implies that the fixed point Hamiltonian for the BKT transition point has the Luttinger parameter $K=2$ instead of $K=1/2$.
It appears that the effective field theory in this case no longer has the SU(2) symmetry and the Level Spectroscopy may not apply.
However, in Ref.~\cite{nomura1998symmetry}, for Level Spectroscopy of the corresponding class of quantum spin chains, ``half-vortex operators'' $V_{\pm 1/2}$ were introduced by twisting the boundary condition.
At the BKT transition, one of the half-vortex states $V^s_{1/2}=\sqrt{2}\sin{\phi}$ becomes degenerate with the spin-wave states $W_{\pm 2}=e^{\pm 2i \theta}$ to form an SU(2) triplet, thereby enabling the application of the Level Spectroscopy.
Thus, in order to apply the Level Spectroscopy to the 2D classical XY model, we need to calculate the spectrum of the transfer matrix (which corresponds to the energy levels of 1D quantum Hamiltonian), under the periodic and twisted boundary conditions.

\section{Level Spectroscopy with Tensor Network Renormalization}
Now we demonstrate a successful implementation of Level Spectroscopy based on TNR for the XY model in Eq.~\eqref{definition_XY}.
First, the partition function of the model is represented in terms of a tensor network, using a series expansion~\cite{PhysRevD.88.056005,PhysRevE.100.062136}
\begin{align}
e^{\beta\cos(\theta_i-\theta_j)}=\sum_{n=-\infty}^{\infty}e^{in(\theta_i-\theta_j)}I_n(\beta),\label{Fourier_cos}
\end{align}
where $I_n$ is the modified Bessel functions of the first kind.
In the practical calculations, we cut off the sum at $|n|=15$ which is verified as sufficient.

In the tensor network representation of the XY model on the square lattice, the tensor on each site has four ``legs'' corresponding to the interactions with the four nearest neighbors. If we contract the legs of $L$ horizontally aligned tensors with the periodic boundary condition, the remaining tensor is nothing but the transfer matrix in the vertical direction for the system of width $L$. As the transfer matrix corresponds to the Hamiltonian of the 1D quantum system, the Level Spectroscopy can be applied to its spectrum. 
In practice, the transfer matrix is obtained by contracting horizontal legs of a single tensor obtained after $N$ steps of TNR.
Since each step of TNR corresponds to rescaling of the lattice spacing by $\sqrt{2}$, this procedure gives the transfer matrix for the system of width $L=\sqrt{2}^N$.
The eigenvalues $\lambda_n(L)$ thus obtained are related to the energy spectrum $E_n(L)$ of the corresponding 1D quantum Hamiltonian as 
\begin{align}
\frac{\lambda_n(L)}{\lambda_0(L)}=\exp(-2\pi x_n(L)),
\end{align}
where we define the rescaled energy levels $x_n$ by $E_n(L)-E_0(L) =2\pi x_n(L) /L$~\footnote{The eigenspectrum of the transfer matrix is $\lambda_n=a e^{-LE_n(L)}$ with a constant prefactor $a$ This is a natural extension of a standard TNR technique\cite{PhysRevB.80.155131}.}.
In this way, we can read off $x_{W_{\pm 2}}$. On the other hand, $x_{V^s_{1/2}}$ appears in the spectrum of the transfer matrix with the twisted boundary condition. 
By introducing a twist of angle $2\pi/L$ in the original model as
\begin{align}
H=-\sum_{\langle{x}\rangle}\cos(\theta_i-\theta_j-\frac{\pi}{L})-\sum_{\langle{y}\rangle}\cos(\theta_i-\theta_j) ,
\label{XY twist}
\end{align}
we can also obtain the transfer matrix spectrum for width $L$ with the twisted boundary condition. In particular, $x_{W_{\pm 2}}$ and $x_{V^s_{1/2}}$ appear as the fourth/fifth(degenerate) and the leading eigenvalues in the transfer matrix in the periodic and twisted boundary conditions, respectively in the vicinity of $T_c$. The scaling dimensions of the twisted boundary condition is measured against the leading eigenvalue of the periodic boundary condition.
\begin{figure*}[tb]
    \centering
    \includegraphics[width=178mm]{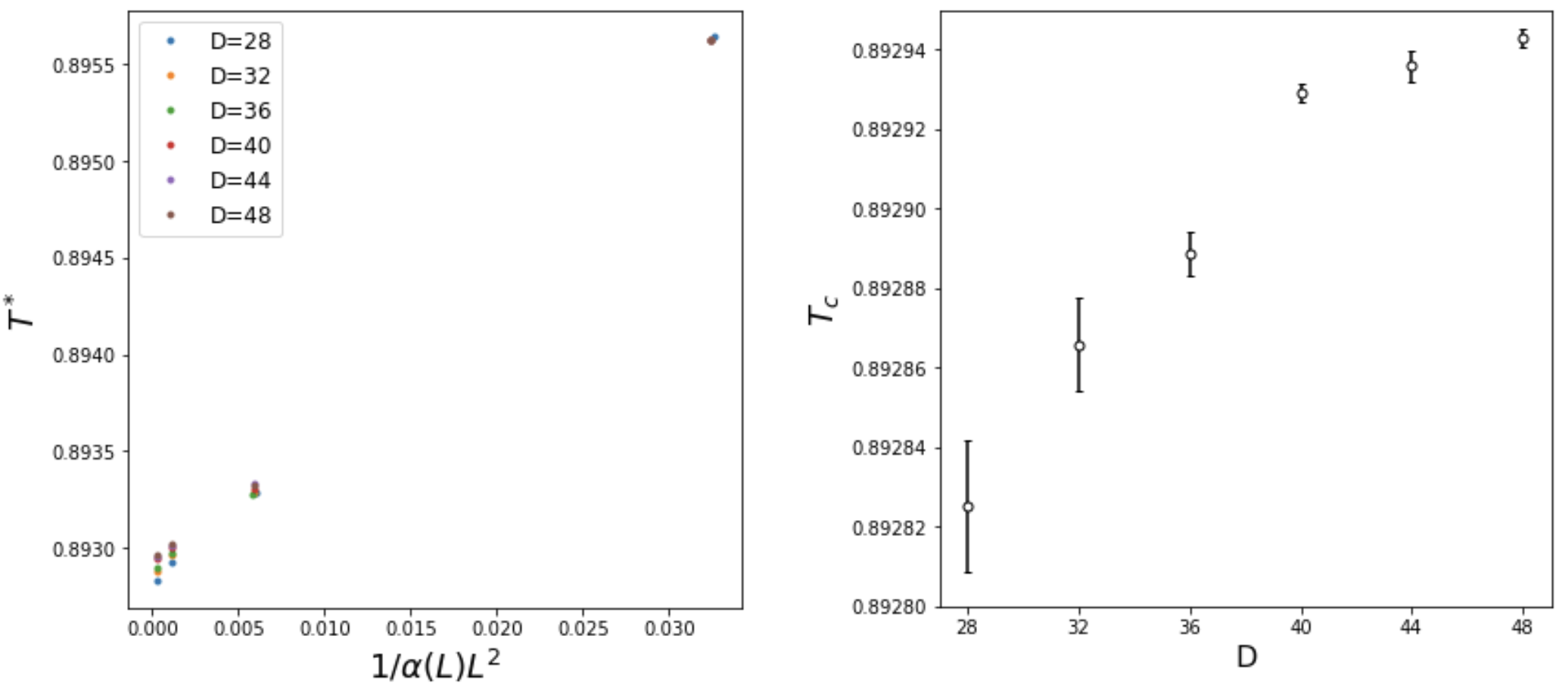}
    \caption{The level-crossing temperature $T^*$ plotted against $\frac{1}{\alpha(L)L^2}$ at $L=4,8,16$ and 32. The data from the bond dimension $D=28,32,36,40,44$ and 48 are shown. The right panel shows the estimated $T_c$ by linear fitting of $L=8,16$ and 32 and its error in the fitting is shown as well. Note that the discrepancy between the estimates of $T_c$ with $D=28$ and $D=48$ is only about $10^{-4}$, which is comparable to the errors in the best existing numerical estimates (see Table~\ref{previous_XY_result}).}
    \label{tstar_extrapolation}
\end{figure*}
\par{}
Let $x_{W_2}$ and $x_{V^s_{1/2}}$ be the rescaled energy levels corresponding to the SU(2) triplet.
While $x_{\calO}$ is given by the scaling dimension of the operator $\calO$ in a pure CFT without any perturbation, it receives corrections from the irrelevant perturbations ($y_\calK, y_V$ and $H'$) and depends on the system size $L$.
Up to the first order in $t$ and $H'$ (but to all orders in $g$), and using $t \propto T-T_c$, we find
\begin{eqnarray}
x_{W_2}(L) &=& x_c(g) + \alpha_W(L) (T-T_c) + \delta_W(L)\nonumber\\
x_{V^s_{1/2}}(L) &=& x_c(g) + \alpha_V(L) (T-T_c) +\delta_V(L)\nonumber 
\end{eqnarray}
where we employed the degeneracy (effective SU(2) symmetry) $x_{W_2}={x_{V^s_{1/2}}}=x_c(g)$ on the BKT transition line $t=0$.
$\delta_W$ and $\delta_V$ represent the first order corrections due to the irrelevant perturbation $H'$.
In the Level Spectroscopy, the transition temperature is first estimated by the crossing point $T^*$
between the two levels $x_{W_2}$ and $x_{V^s_{1/2}}$.
As far as the irrelevant perturbation $H'$ is ignored, it gives the exact transition temperature which corresponds to $t=0$, removing
the notorious logarithmic corrections to the all order.
However, because of the irrelevant perturbation $H'$, the crossing point $x_{W_2} = x_{V^s_{1/2}}$ rather gives 
\begin{equation}
    T^* = T_c - \frac{\delta_W(L) - \delta_V(L)}{\alpha_W(L) - \alpha_V(L)} .
\end{equation}
$\alpha_W(L)$ and $\alpha_V(L)$ can be extracted as the slope of the corresponding levels in Fig.~\ref{level_crossing} for each size $L$.
Since the leading irrelevant perturbations (other than the marginally irrelevant $\calK$ and $V$) to the fixed point Hamiltonian are 
the square of the stress-energy tensor with the scaling dimension $4$, the standard CFT analysis~\cite{cardy1984conformal} implies
$\delta_W \propto \delta_V \propto 1/L^2$.
Thus, the level-crossing point $T^*(L)$ between $x_{W_2}$ and $x_{V_{1/2}}^{s}$ obtained for the system size $L$ is related to the true transition temperature $T_c$ as
\begin{align}
T^*(L)=T_c- \text{const.} \frac{1}{\alpha(L) L^2} ,
\label{ShiftT}
\end{align}
where $\alpha(L) = \alpha_W(L)-\alpha_V(L)$. 
Therefore, the transition point can be extrapolated linearly by plotting $T^*$ against $\frac{1}{\alpha(L)L^2}$.
The original Level-spectroscopy implicitly assumed that $\alpha(L)$ is independent of $L$. This was a reasonable assumption because the system size employed there only ranged from $L=4$ to $12$. However, in our study, the system size is extended up to $L=32$ and the size dependence of $\alpha(L)$ is not negligible. Indeed, in our numerical estimates of $\alpha(L)$, we found a weak dependence on $L$, which is consistent with the asymptotic behavior $\alpha(L) \sim \log{L}$ determined by the RG equation~\eqref{RG_BKT}.
The left panel of Fig.~\ref{tstar_extrapolation} exhibits the size dependence of the level-crossing temperature $T^*$. The data show a good agreement with the theoretical prediction.
The data for $L=4$, which appears off from Eq.~\eqref{ShiftT}, are presumably affected by more irrelevant perturbations with the scaling dimension $6$ and higher. Thus we carry out the linear extrapolation from the data for $L=8,16, 32$ where Eq.~\eqref{ShiftT} holds almost perfectly to obtain $T_c$.
On the other hand, any practical calculation based on a tensor network is inherently limited by a finite bond dimension.
The TNR is also accurate for the system sizes only up to the maximum correlation length imposed by the finite bond dimension.
In the present Level Spectroscopy approach, however, we can obtain very accurate results from only moderately large systems up to $L=32$, where the TNR with the bond dimension $D=48$ is sufficient~\footnote{We confirmed numerically that the effective correlation length $\xi_D$ at the transition point is $\xi_D\simeq0.3D^\kappa$, where $\kappa$ coincides with the exponents in Ref.~\cite{pollmann2009theory}. The effective correlation lengths are $\xi_D\sim 26,54$ for $D=28,48$ respectively. The relatively large error bar for $D=28$ appears because the correlation length is smaller than the system size at $L=32$. On the other hand, since the effective correlation is long enough for $D=48$, the data points correctly capture $\sim 1/\alpha(L)L^2$, leading to a small error bar. More details on the effect of the finite bond dimension will be discussed in a separate publication~\cite{Ueda-inprep}.}.
This is manifest in the right panel of Fig.~\ref{tstar_extrapolation}, which shows the extracted $T_c$ as a function of the bond dimension $D$: 
the dependence on $D$ saturates at $D \sim 40$, and the BKT transition temperature is identified as $T_c=0.892943(2)$ at $D=48$.
As shown on Table~\ref{previous_XY_result}, our result has a higher precision than previous studies by an order of magnitude,
\begin{table}[b]
\begin{ruledtabular}
\begin{tabular}{c|cl}
Monte Carlo(1979)\cite{PhysRevB.20.3761}      & 0.89    \\ \hline
Monte Carlo(2005)\cite{Hasenbusch_2005}      & 0.8929(1)  \\ \hline
Monte Carlo(2012)\cite{komura2012large}      & 0.89289(6) \\ \hline
Monte Carlo(2013)\cite{Hsieh_2013}      & 0.8935(1)  \\ \hline
Series expansion(2009)\cite{PhysRevE.79.011107} & 0.89286(8) \\ \hline
HOTRG(2014)\cite{PhysRevE.89.013308}             & 0.8921(19)  \\ \hline
VUMPS(2019)\cite{PhysRevE.100.062136}         & 0.8930(1)  \\ \hline
HOTRG(2020)\cite{Jha_2020}                 &0.89290(5) \\ \hline
present work                & 0.892943(2) \\
\end{tabular}
\end{ruledtabular}
\caption{Comparison of the estimated critical temperature of the 2D classical XY model.\label{previous_XY_result}}
\end{table}
\section{RG flow of the XY model}
\begin{figure}[t]
    \centering
    \includegraphics[width=86mm]{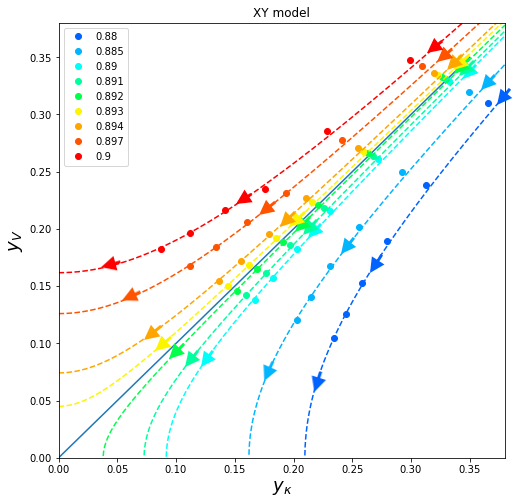}
    \caption{The numerically computed RG flow of the XY model. The data was extracted from $T=0.88\sim0.9$ and $L=$16, 32, 64, 128, 256 and 512. The dotted lines are rough fittings of the plots by hyperbolas and the blue line indicates the BKT line $y_\calK=y_V$.}
    \label{visualization_RG}
\end{figure}

Finally, one can extract the coupling constants of the sine-Gordon model to visualize the RG flow.
Ignoring $H'$ here, up to the second order in $y_\calK$ and $y_V$, the lowest rescaled energy levels are given as
\begin{align}
x_{W_{\pm2}}&=\frac{1}{2}-\frac{y_\mathcal{K}}{4}+\frac{1}{4}y_V^2,\\
x_{V^s_{1/2}}&=\frac{1}{2}+\frac{y_\calK}{4}-\frac{y_V}{2}+\frac{1}
{8}(y_\mathcal{K}^2+2y_\calK y_V-y_V^2),\\
x_{V^c_{1/2}}&=\frac{1}{2}+\frac{y_\calK}{4}+\frac{y_V}{2}+\frac{1}
{8}(y_\mathcal{K}^2-2y_\calK y_V-y_V^2),
\end{align}
where $V^c_{1/2}=\sqrt{2}\cos\phi$ is the remaining singlet.
We read off $x_{W_{\pm2}}$ from Ref~\cite{lukyanov1998low} and determine $x_{V^s_{1/2}}$ by imposing a restriction by the symmetry. (See the Appendix) Then, the running coupling constants are extracted from the numerically obtained energy levels as
\begin{equation}
\begin{aligned}
    y_\calK & \sim 2-4x_{W_{\pm  2}}+(x_{V^c_{1/2}}-x_{V^s_{1/2}})^2,
    \\
    y_V & \sim (x_{V^c_{1/2}}-x_{V^s_{1/2}})/(1-\frac{1}{2}y_\calK),
\end{aligned}
\label{eq:y_in_x}
\end{equation}
which are valid up to $O(y^3)$.
Fig.~\ref{visualization_RG} shows the obtained RG flow of the XY model in the vicinity of the transition temperature. The theoretical trajectories of Eq.~\eqref{RG_BKT} are hyperbolas $y_V^2-y_\calK^2=$const. and our result matches perfectly with it: The low-temperature phase shown with blue colors flows to the $y_V=0$ critical line, whereas the high-temperature region does not. In the middle of two phases are the RG flow of the critical temperature $T_c=0.893$ marked with yellow dots, which is on top of the blue BKT line.
This estimate of $T_c$ is consistent with our earlier estimate shown in
Table~\ref{previous_XY_result}~\footnote{Our estimate of $T_c$ in Table~\ref{previous_XY_result} is more accurate, because of the following reasons.
First, here we deal with large systems up to $L=512$ which is larger than the effective correlation length imposed by the finite bond dimension.
The results, however, still seem to be accurate enough for the present purpose of tracking the RG flow (see Fig.~\ref{tstar_extrapolation}).
Secondly, the transition point in Table~\ref{previous_XY_result} determined by the level crossing is valid in all orders of the marginal coupling $g$ while Eq.~\eqref{eq:y_in_x} is valid only in $O(y^3)$.
Furthermore, the effects of the irrelevant perturbation $H'$ are removed as in Fig.~\ref{tstar_extrapolation} to obtain the value in Table~\ref{previous_XY_result}.}.
\section{Conclusion}
We analyzed the eigenspectrum of the renormalized tensors for the classical 2D XY model at finite renormalization steps,
in terms of finite-size scaling of CFT.
The BKT transition is described in terms of two marginal couplings $y_\calK$ (spin wave stiffness) and $y_V$ (vortex fugacity), and the transition point can be identified with $y_\calK=y_V$ where a hidden SU(2) symmetry emerges.
By exploiting the SU(2) symmetry, we determine the transition temperature with a record precision from the spectrum, improving the Level Spectroscopy method developed for 1D quantum systems.
Furthermore, we note that SU(2) symmetry is no longer a requisite by regarding Level Spectroscopy as \emph{determination of transition temperatures based on the coupling constants of the underlying field theory extracted from the finite-size spectrum.} One can extract the running coupling constants from a finite-size spectrum by following Eqs.~(\ref{perturbation_CFT}, \ref{fss_scaling_dimension}, \ref{finite_lambda_x}) in the Appendix, and tracking the RG flow exhibits the phase separation visually similar to Fig.~\ref{visualization_RG}.
Our method is advantageous because finite-size effects are substantially reduced by the removal of the logarithmic corrections and the effects of the finite bond-dimension are suppressed at the moderate system sizes needed for our analysis. 
We also numerically tracked the evolution of the two coupling constants $y_\calK$ and $y_V$ as the system size is increased near the transition to visualize the celebrated Kosterlitz RG flow.
The present work connects the conceptual understanding of one of the most important examples of phase transitions to the contemporary numerical algorithm, paving the way to further developments.
\section*{Acknowledgements}
We thank Laurens Vanderstraeten, Frank Verstraete, Hosho Katsura, Kiyohide Nomura, and Kiyomi Okamoto for stimulating discussions.
This work was supported in part by MEXT/JSPS KAKENHI Grant Nos. JP17H06462 and JP19H01808, and JST CREST Grant No. JPMJCR19T2.
A part of the computation in this work has been done
using the facilities of the Supercomputer Center, the Institute for Solid State Physics, the University of Tokyo.
\appendix
\section{Review on Level Spectroscopy}
\subsection{Overview}
The original Level Spectroscopy was first developed to determine the transition temperature of the BKT transition for quantum spin chains. In particular, they investigated the effective Hamiltonian for the spin 1/2 XXZ spin chain. It has a phase transition (dimer to CDW) and the gapless phase is described by the Tomonaga-Luttinger liquid (TLL). The Lagrangian of the gapless phase is essentially 
\begin{align}
\mathcal{L}=\frac{1}{2\pi K}\left(\partial_\mu\phi\partial^\mu\phi\right),\label{TLL_lagrangian}
\end{align}
where $K$ is called Luttinger parameter. The bosonic field $\phi$ is compactified in the bulk as 

    $$\phi\sim\phi+\pi$$
The canonical quantization of the field theory Eq.~\eqref{TLL_lagrangian} is done by requiring the canonical conjugate field $$\Pi\equiv\frac{\delta\calL}{\delta\dot{\phi}}=\frac{1}{\pi K}\dot{\phi},$$
where $\dot{\phi}=\partial_t\phi$. The resulting Hamiltonian is 
\begin{align}
 H^{\text{TLL}} = \int dx\left[\frac{\pi K}{2}\Pi^2+\frac{1}{2\pi K}(\partial_x \phi)^2\right].
 \label{Gaussian}
\end{align}
Following the equation of motion of the Lagrangian Eq.~\eqref{TLL_lagrangian}
$$\partial_\mu\partial^\mu\phi=0,$$
we can decompose the field $\phi$ into the right- and left-mover as
$$\phi(t,x)=\varphi(x-t)+\bar{\varphi}(x+t).$$
The dual field $\theta$ is then defined as 
$$\theta(t,x)=\frac{1}{K}\left(\varphi(x-t)-\bar{\varphi}(x+t)\right).$$
$\theta$ is also compactified as
$$\theta\sim\theta+2\pi.$$
In physical applications, often $\theta$ represents the angular variable corresponding to the microscopic $U(1)$ symmetry, and it is indeed the continuous counterpart of the lattice variable $\theta_i$ of the classical XY model in Eq.~(1) of the main text.
(In the next section, we apply Wick rotation as $\tau=it,~ z=x+i\tau,~$and $\bar{z}=x-i\tau$ so as to consider the CFT on the plane.)
\par{}Since the filed theory Eq.~\eqref{TLL_lagrangian} is scale-invariant, it describes the critical and gapless phases of 2D classical systems and 1D quantum phases, respectively.
However, the lattice models, generally speaking, contain perturbations at  finite sizes and the XXZ spin chain is no exception. The field theoretical description of the system is the sine-Gordon model in the long wave-length limit near the critical parameters as,
\begin{align}
    H_{XXZ}=H^{TLL}_{K=1/2}+\int\frac{dx}{2\pi}\left\{\frac{y_\calK}{2} \calK+{y_V}\cos(4\phi)\right\},\label{XXZ_Hamiltonian_appendix}
\end{align}
where $\calK=-2\partial_\mu\phi\partial^\mu\phi=J\bar{J}$ in Table~\ref{notation}. The first and second terms in the perturbations represent the renormalization of $K$ and the creation of vortex-antivortex pairs, respectively. Kosterlitz derived the RG equations, and the transition occurs when $g=y_\calK=y_V$.
On the BKT line $y_\calK=y_V$, 
the system restores SU(2) symmetry~\cite{PhysRevD.12.1684,BANKS1976119} and the SU(2) triplet $V_1^s=\sin(2\phi)$ and $W_{\pm 1}=e^{\pm i\theta}$ becomes degenerate to the higher order loops of of $g$.

The TLL is an example of Conformal Field Theory (CFT) in $1+1$ dimensions.
Cardy~\cite{cardy1984conformal} showed that,
in an unperturbed CFT, the energy eigenvalues $E_n$ of the primary states in a quantum Hamiltonian of length $L$ is given as
\begin{equation}
 E_n(L) - E_0(L) = \frac{2 \pi}{L} x_n ,    
\end{equation}
where $E_0$ is the ground state energy and $x_n$ is the scaling dimension of the corresponding primary operator.
Furthermore, perturbations to the CFT Hamiltonian give corrections to this relation~\cite{cardy1986operator}.
Conversely, by analyzing the finite-size energy spectrum, one can estimate the perturbations to the CFT.
This is the foundation of the Level Spectroscopy method.

\subsection{Derivation of Level Spectroscopy}

In the original papers~\cite{nomura1995correlation,OKAMOTO1992433,nomura1994critical,nomura1998symmetry,matsuo2006berezinskii} on Level Spectroscopy, the degeneracy of $V_1^s$ and $W_{\pm1}$ was discussed in terms of the correspondence between the sine-Gordon model and the SU(2) Thirring model~\cite{PhysRevD.12.1684,BANKS1976119}. Here, we present the derivation of Level Spectroscopy in more intuitive manner.
\par{} The comformal weights of $:e^{im\theta+i2n\phi}:$ are $h_{m,n}=\frac{1}{2}(\frac{m}{2\sqrt{K}}+\sqrt{K}n)^2$ and $\bar{h}_{m,n}=\frac{1}{2}(\frac{m}{2\sqrt{K}}-\sqrt{K}n)^2$. The scaling dimension is defined as a sum of conformal weights as $x_{m,n}=\frac{m^2}{4K}+n^2K$ as in the main text. At $K=1/2$, $(m,n)=(\pm1,\pm1)$ and $(m,n)=(\pm1,\mp1)$ become currents respectively and the Hamiltonian gets equivalent to $\rm SU(2)_1$ WZW model. The currents respect the SU(2) current algebra as
\begin{align}
    J^i(z)J^j(w)=\frac{k/2}{(z-w)^2}\delta_{ij}+\frac{i\epsilon_{ijl}J^l(w)}{z-w},\label{current_algebra}
\end{align}
where $k=1$ and $i=0,\ 1,\ 2$(or equivalently $i=z,\ x,\ y$) in the current case.  The notations are listed in Table~\ref{notation} and one can easily check that they follow Eq.~\eqref{current_algebra}. Since the algebra is a representation of SU(2), it can be interpreted as a spin system. For instance, $\ket{\uparrow}^L$ is $e^{i2\varphi}\ket{0}$ because its SU(2) charge is 
\begin{align}
    \oint_0\frac{dz}{2\pi i}J^0(z)\ket{\uparrow}^L&=\oint_0\frac{dz}{2\pi i}\frac{1/2}{z}e^{i2\varphi(0)}\ket{0}\nonumber\\
    &=\frac{1}{2}\ket{\uparrow}^L,
\end{align}
whereas $J^+\ket{\uparrow}=0$ because it has no pole as $J^+(z)e^{i2\varphi(w)}\sim (z-w)e^{i(4\varphi(z)+2\varphi(w))} $.
Using these notations, Eq.~\eqref{XXZ_Hamiltonian_appendix} at $g=y_\calK=y_V$ can be re-written as
\begin{align}
    H_{XXZ}&=H^{WZW}_{k=1}-\int\frac{dx}{2\pi}g\left\{J^0\bar{J}^0+\frac{1}{2}(J^+\bar{J}^-+J^-\bar{J}^+)\right\}\nonumber\\
    &=H^{WZW}_{k=1}-\int\frac{dx}{2\pi}g(\bm{J^L}\cdot \bm{J^R}),
\end{align}
where $H^{WZW}_{k=1}$ is a sum of the Sugawara energy-momentum tensors as $H^{WZW}_{k=1}=\int T(z)+\bar{T}(\bar{z})=\int\frac{1}{3}[(\bm{ J^L})^2+(\bm{J^R})^2]$~\cite{francesco2012conformal}.
 The lowest lying eigenstates of $\int\frac{dx}{2\pi}(\bm{J^L}\cdot \bm{J^R})$ are the spin singlet and triplets with their eigenvalues $-\frac{3}{4}$ and $\frac{1}{4}$. Using Table~\ref{notation}, the primary operators corresponding these states are
\begin{align}
    \frac{1}{\sqrt{2}}(\ket{\uparrow\downarrow}-\ket{\downarrow\uparrow})&=\lim_{z,\bar{z}\rightarrow0}\sqrt{2}\cos\left(2\phi(z,\bar{z})\right)\ket{0},\nonumber\\
    \ket{\uparrow\uparrow}&=\lim_{z,\bar{z}\rightarrow0}-e^{i\theta(z,\bar{z})}\ket{0},\\
    \frac{1}{\sqrt{2}}(\ket{\uparrow\downarrow}+\ket{\downarrow\uparrow})&=\lim_{z,\bar{z}\rightarrow0}i\sqrt{2}\sin\left(2\phi(z,\bar{z})\right)\ket{0},\nonumber\\
    \ket{\downarrow\downarrow}&=\lim_{z,\bar{z}\rightarrow0}e^{-i\theta(z,\bar{z})}\ket{0}.
\end{align}
As $\bm{J^+}$ commutes with $\bm{J^2}$, $V_1^s$ and $W_{\pm1}$ are degenerate. In particular, the energy levels of the singlet and triplets to the first order can be evaluated as
\begin{align}
    E_{singlet}&=\frac{2\pi}{L}\left(\frac{1}{4}+\frac{1}{4}-g\frac{1}{2}(0-2\times\frac{1}{2}(1+\frac{1}{2})\right)\nonumber\\
    &=\frac{2\pi}{L}(\frac{1}{2}+\frac{3}{4}g),\\
     E_{triplets}&=\frac{2\pi}{L}\left(\frac{1}{4}+\frac{1}{4}-g\frac{1}{2}(1(1+1)-2\times\frac{1}{2}(1+\frac{1}{2})\right)\nonumber\\
     &=\frac{2\pi}{L}(\frac{1}{2}-\frac{1}{4}g).
\end{align}
Thus, the scaling dimensions of the triplets are $\frac{1}{2}-\frac{1}{4}g+O(g^2)$. 
\begin{table*}[t]
\begin{ruledtabular}
\begin{tabular}{||c|c||c|c||c|c||}
\hline
Boson field               &  Definition                                 & Currents ($k=1$ WZW) &   Corresponding operator                           & The highest weight   &     Corresponding state                           \\ \hline
$\phi\sim\phi+\pi$      & $\varphi(z)+\bar{\varphi}(\bar{z})$           & $J^0$                & $\frac{1}{\sqrt{2}}J$        & $\ket{\uparrow}^L$   & $\lim_{z\rightarrow0}e^{i2\varphi(z)}\ket{0}$         \\ \hline
$\theta\sim\theta+2\pi$ & $2(\varphi-\bar{\varphi})$        & $\bar{J}^0$          & $-\frac{1}{\sqrt{2}}\bar{J}$ & $\ket{\uparrow}^R$   & $\lim_{\bar{z}\rightarrow0}-e^{-i2\bar{\varphi}(\bar{z})}\ket{0}$ \\ \hline
$J$                     & $i2\sqrt{2}\partial\varphi$       & $J^{\pm}=J^1\pm i J^2$            & $e^{\pm i4\varphi}$             & $\ket{\downarrow}^L$ & $\lim_{z\rightarrow0}e^{-i2\varphi(z)}\ket{0}$        \\ \hline
$\bar{J}$               & $i2\sqrt{2}\partial\bar{\varphi}$ & $\bar{J}^{\pm}=\bar{J}^1\pm i\bar{J}^2$      & $-e^{\mp i4\bar{\varphi}}$      & $\ket{\downarrow}^R$ & $\lim_{\bar{z}\rightarrow0}e^{i2\bar{\varphi}(\bar{z})}\ket{0}$   \\ \hline
\end{tabular}
\caption{The notation and normalized currents.\label{notation}}
\end{ruledtabular}
\end{table*}

\subsection{Implementation of Level Spectroscopy}
For small system sizes, there are also irrelevant operators that do not respect SU(2) symmetry due to the lattice anisotropy. The leading irrelevant ones are $T^2, \bar{T}^2, T\bar{T}$, where $T$ and $\bar{T}$ are holomorphic and antiholomorphic components of the energy-momentum tensor (not to be confused with the temperature $T$).
Since they have scaling dimension $4$,
it splits the energy levels of the triplets by $\sim 1/L^2$.
While $W_{\pm1}$ remain degenerate, $V_1^s$ is no longer at the same energy level. Nonetheless, the OPE coefficients with these irrelevant operators must be almost the same (conformal weights are almost the same). Thus, the level-crossing point is closed to the true transition point even at small $L$.

In the case of the 2D classical XY model, as discussed in the main text, the BKT transition is described by making the replacements
 on that for the $S=1/2$ XXZ chain: $2\phi\rightarrow\phi,\quad\theta\rightarrow2\theta$, and $K=1/2\rightarrow K=2$.
In this context, $W_{\pm2}$ and $V_{1/2}^s$ should be degenerate and form a SU(2) triplet, where $V_{1/2}^s$ is one of the linear combinations of $V_{\pm 1/2}$ corresponding to the insertion of the $\pi$-twist operator (antiperiodic boundary condition).

\section{Calculation of $y_\calK$ and $y_V$}

\subsection{Lukyanov's result}
Combining the exact Bethe ansatz solution and the low-energy effective field theory,
Lukyanov~\cite{lukyanov1998low} studied the``vacuum'' (ground-state) energy of the XXZ chain for a given total magnetization under the twisted boundary condition with the twist parameter $2 \pi \theta$. (Here we introduce $\theta$ as the twist parametrizing the boundary condition as in Ref.~\cite{lukyanov1998low}. This should not be confused with the use of $\theta$ as a field variable elsewhere in this paper.) 
His remarkable results have been verified by several numerical studies.
While the results are limited to $\theta <1$ in Ref.~\cite{lukyanov1998low}, we may access the excited states corresponding to $V_{\pm 1}$ under the untwisted (periodic) boundary condition.
However, in this limit $\theta=1$, the mixing between $V_{\pm 1}$ states by the allowed vortex perturbation $V^c_2$ must be taken into account, as it was indeed suggested by Lukyanov~\cite{lukyanov1998low}.
Below we will demonstrate that the mixing is necessary to reproduce the split between triplet and singlet levels at the BKT transition, which is required by the emergent SU(2) symmetry as discussed in the main text.

Lukyanov's results can be also translated into our problem of the classical 2D XY model by the simple replacement $2\phi \to \phi$ as discussed above.
In the following, we present an analysis in the context of the classical 2D XY model.
According to Lukyanov, the scaling dimension to the third order is
\begin{align}
x_{m,n}&=\frac{m^2}{8}(1-\frac{y_\mathcal{K}}{2}+\frac{y_V^2}{4}-\frac{7}{32}y_\mathcal{K}y_V^2)\nonumber\\
&+\frac{|m|}{32}(2y_V^2-y_\mathcal{K}y_V^2)\nonumber\\
&+2n^2(1+\frac{y_\mathcal{K}}{2}+\frac{y_\mathcal{K}^2}{4}+\frac{y_\mathcal{K}^3}{8}-\frac{y_V^2}{4}-\frac{y_\mathcal{K}y_V^2}{32}),
\end{align}
where $x_{m,n}$ is the scaling dimension of $:e^{im\theta+i2n\phi}:$. In particular, 
\begin{align}
x_{1,0}&=\frac{1}{8}-\frac{y_\mathcal{K}}{16}+(\frac{3}{32}-\frac{15}{256}y_\mathcal{K})y_V^2,\\
x_{2,0}&=\frac{1}{2}-\frac{y_\mathcal{K}}{4}+(\frac{1}{4}-\frac{11}{64}y_\mathcal{K})y_V^2\label{lukyanov_W2},\\
x_{0,1/2}&=\frac{1}{2}+\frac{y_\mathcal{K}}{4}+\frac{1}
{8}(y_\mathcal{K}^2-y_V^2)+\frac{y_\mathcal{K}^3}{16}-\frac{1}{64}y_\mathcal{K}y_V^2\label{lukyanov_12}.
\end{align}
In Eq.~\eqref{lukyanov_12}, however, the energy repulsion between $x^s_{0,1/2}$ and $x^c_{0,1/2}$, due to the cosine term, is not considered here. Hence, we shall determine it in order to calculate $y_\calK$ and $y_V$ to the second-order.

\subsection{Calculation of $x_{V^s_{1/2}}$}
We deduce the energy level of $V^s_{1/2}$ using two facts:\\
   $\quad\quad\cdot$ $W_{\pm2}$ and $V^s_{1/2}$ are exactly degenerate on the BKT line.\\
   $\quad\quad\cdot$ $x_{V^s_{1/2}}+x_{V^c_{1/2}}=2x_{0,1/2}$, which is Eq.~\eqref{lukyanov_12}.\\
First, from Eq.~\eqref{lukyanov_W2} we find the energy level of the triplets on the BKT line $g=y_\calK=y_V$ is 
\begin{align}
x_{triplet}=\frac{1}{2}-\frac{g}{4}+\frac{g^2}{4}\label{BKT_triplet}.
\end{align}
Let us define $x_{V^c_{1/2}}-x_{V^s_{1/2}}=2b$. Then, using the second fact, we find that Eq.~\eqref{BKT_triplet} is larger than Eq.~\eqref{lukyanov_12} by $b$. Simple calculation leads to the power expansions of $b$ with $g$ as 
\begin{align}
b=\frac{g}{2}-\frac{g^2}{4}.
\end{align}
Given that $b$ should be odd under $y_V\rightarrow -y_V$ the explicit form of $b$ is deduced as
\begin{align}
    b=(1-\frac{1}{2}y_\calK)\frac{y_V}{2}.
\end{align}
The final forms of the singlet and triplets are 
\begin{align}
x_{W_{\pm1}}&=\frac{1}{8}-\frac{y_\calK}{16}+\frac{3}{32}y_V^2,\label{2nd_1}\\
x_{W_{\pm2}}&=\frac{1}{2}-\frac{y_\calK}{4}+\frac{1}{4}y_V^2,\label{2nd_2}\\
x_{V^s_{1/2}}&=\frac{1}{2}+\frac{y_\calK}{4}-\frac{y_V}{2}+\frac{1}{8}(y_\calK^2+2y_\calK y_V-y_V^2),\label{2nd_3}\\
x_{V^c_{1/2}}&=\frac{1}{2}+\frac{y_\calK}{4}+\frac{y_V}{2}+\frac{1}{8}(y_\calK^2-2y_\calK y_V-y_V^2).\label{2nd_4}
\end{align}

\subsection{Perturbative calculation to the first order based on Conformal Field Theory}
Generally speaking, $x_n(L)$ depends on the system size due to the relevant/irrelevant perturbations to the fixed-point Hamiltonian.  Let the Hamiltonian with the system size $L$ be
\begin{align}
 \hat{H}(L)  &= \hat{H}^* + \sum_j g_j \int_0^L dx \; \hat{\Phi}_j(x),\label{perturbation_CFT}
 \end{align}
where $\hat{H}^*$ is the Hamiltonian of the fixed point, $\hat{\Phi}_j(x)$ is an operator with the scaling dimension $x_j$  and $g_j$ is a corresponding coupling constant.
The scaling dimensions of a finite-size system is then described by 
\begin{align}
x_n(L)&=x_n+2\pi\sum_jc_{nnj}g_j\left(\frac{2\pi}{L}\right)^{x_j-2}\nonumber\\
&=x_n+2\pi\sum_jc_{nnj}g_j(L).\label{fss_scaling_dimension}
\end{align}
 $c_{ijk}$ is the operator product expansion(OPE) coefficient~\cite{cardy1984conformal,cardy1986operator}. Following a standard calculation from Tomonaga-Luttinger liquid, we find that the OPE coefficients as $c_{\calK W_{\pm m}W_{\pm m}}=-\frac{m^2}{4K}$, $c_{\calK V_{\pm n}V_{\pm n}}=n^2K$, $c_{V^c_1V^c_{1/2}V^c_{1/2}}=\frac{1}{\sqrt{2}}$ and $c_{V^c_1V^s_{1/2}V^s_{1/2}}=-\frac{1}{\sqrt{2}}$.
 On the other hand, the effective Hamiltonian for the classical XY model is 
 \begin{equation}
 \hat{H}  = \hat{H}^{\mbox{\scriptsize TLL}}_{K=2} + 
\int_0^L dx \; \left\{
\frac{y_\calK}{4\pi} \calK(x)  + \frac{y_V}{2\sqrt{2}\pi} V^c_1(x)
\right\} .
\label{eq.BKT2_gKgV}
\end{equation}
Comparing Eq.~\eqref{eq.BKT2_gKgV} with Eq.~\eqref{perturbation_CFT}, we identify $g_\calK=\frac{y_\calK}{4\pi}$ and $g_V=\frac{y_V}{2\sqrt{2}\pi}$. Substituting them into Eq.~\eqref{fss_scaling_dimension}, we find
\begin{align}
    x_{W_{\pm m}}&=\frac{m^2}{8}+2\pi c_{W_{\pm m}W_{\pm m}\calK}g_\calK\nonumber\\
    &=\frac{m^2}{8}(1-\frac{y_\calK}{2})\\
    x_{V^s_{1/2}}&=\frac{1}{2}+2\pi(c_{V^s_{1/2}V^s_{1/2}\calK}g_\calK+c_{V^s_{1/2}V^s_{1/2}V^c_{1/2}}g_V)\nonumber\\
    &=\frac{1}{2}+\frac{y_\calK}{4}-\frac{y_V}{2}\\
    x_{V^c_{1/2}}&=\frac{1}{2}+2\pi(c_{V^c_{1/2}V^c_{1/2}\calK}g_\calK+c_{V^c_{1/2}V^c_{1/2}V^c_{1/2}}g_V)\nonumber\\
    &=\frac{1}{2}+\frac{y_\calK}{4}+\frac{y_V}{2}
\end{align}
This is in agreement with Eqs.~(\ref{2nd_1}-\ref{2nd_4}) to the first order. The scaling dimension
\begin{align}
x_n(L)=\frac{1}{2\pi}\ln(\lambda_0/\lambda_n)\label{finite_lambda_x}
\end{align}
in the main text changes as we renormalize the tensor(as we change the system sizes $L$), and so are $y_\calK(L)$ and $y_V(L)$. Therefore, we can compute $y_\calK$ at each scale by measuring the deviation of $x_{W_{\pm 2}}$ from $\frac{1}{2}$ for example. The identification of the relevant energy levels can be done by comparing the exact values of the scaling dimensions in the UV/IR fixed point.
\par{} This approach is applicable as long as the Hamiltonian is in the vicinity of the fixed-point. Combined with TNR, our approach would potentially become a powerful method to quantitatively compute running coupling constants of the field theory from lattice models near the fixed points.
\bibliography{manuscript}

\end{document}